\documentstyle[sprocl2,epsfig,amssymb]{article}
\begin{document}
\thispagestyle{empty}
\begin{flushright}
\large
DTP/98/34 \\ May 1998
\end{flushright}
\vspace{0.65cm}

\begin{center}
\LARGE
{\bf NLO Corrections to Heavy Quark}

\vspace{0.1cm}
{\bf Production with Polarized Beams}

\vspace{1.2cm}
\Large
M.\ Stratmann\\
\vspace{0.5cm}
\large
Department of Physics, University of Durham,\\ 
\vspace{0.1cm}
Durham DH1 3LE, England\\
\vspace{2.4cm}
{\bf Abstract} 
\end{center}
\vspace*{0.5cm}

\noindent
We present a calculation of the NLO QCD corrections 
to heavy flavor photoproduction with longitudinally polarized beams. We apply
our results to study the spin asymmetry for total charm quark production 
which will be used for a first direct determination of $\Delta g$
by the COMPASS experiment. We also briefly discuss the main 
theoretical uncertainties inherent in this calculation.
\normalsize

\vspace{4.0cm}
\noindent
{\it Talk presented at the 6th International Workshop on
'Deep Inelastic Scattering and QCD' (DIS '98), Brussels, Belgium,
April 4-8, 1998.}
\vfill

\setcounter{page}{0}
\newpage
\font\eightrm=cmr8

\bibliographystyle{unsrt} 

\arraycolsep1.5pt

\def\Journal#1#2#3#4{{#1} {\bf #2}, #3 (#4)}

\def\NPB{{\em Nucl. Phys.} B}
\def\PLB{{\em Phys. Lett.}  B}
\def\PRL{\em Phys. Rev. Lett.}
\def\PRD{{\em Phys. Rev.} D}
\def\ZPC{{\em Z. Phys.} C}


\title{NLO CORRECTIONS TO HEAVY QUARK PRODUCTION WITH POLARIZED BEAMS}

\author{MARCO STRATMANN}

\address{Department of Physics, University of Durham, Durham DH1 3LE,
England}

\maketitle\abstracts{We present a calculation of the NLO QCD corrections 
to heavy flavor photoproduction with longitudinally polarized beams. We apply
our results to study the spin asymmetry for total charm quark production 
which will be used for a first direct determination of $\Delta g$
by the COMPASS experiment. We also briefly discuss the main 
theoretical uncertainties inherent in this calculation.}

\section{Introduction}

Despite significant progress in the field of spin-dependent DIS, 
the polarized gluon density $\Delta g$ remains almost 
completely unconstrained \cite{grsv,gs,pdfs} by presently available DIS data.
An important role plays here the lack of any direct constraint 
on $\Delta g$ from other processes. 
Upcoming spin experiments will thus put a special emphasis on 
exclusive measurements to provide further invaluable information for 
a more restrictive analysis of polarized parton densities in the future.
In this context heavy quark $(Q=c,\,b)$ photoproduction is considered 
to be one of the best options to pin down $\Delta g$ because in LO only 
the photon-gluon fusion (PGF) process \cite{lohq} 
(an arrow denotes a longitudinally polarized particle)
\begin{equation}
\label{eq:lopgf}
\vec{\gamma} \vec{g} \rightarrow Q \bar{Q}
\end{equation}
contributes. Since LO estimates of (\ref{eq:lopgf}) are rather unreliable, 
and since we already know from the unpolarized NLO calculation 
\cite{svn,ellis} that the corrections are sizeable and that the 
`clean picture' of (\ref{eq:lopgf}) is obscured by new,
 light quark induced NLO subprocesses, 
the knowledge of the polarized NLO corrections
\cite{bs} is mandatory for a meaningful extraction of $\Delta g$.
In what follows we will briefly highlight the main steps and results
of our calculation \cite{bs}.

\section{Technical Framework}

The NLO QCD corrections to the PGF mechanism in (\ref{eq:lopgf})
consist of three parts: (i) the one-loop virtual corrections,
(ii) the real corrections $\vec{\gamma} \vec{g} \rightarrow Q \bar{Q} g$,
and (iii) a new genuine NLO production mechanism  
$\vec{\gamma} \vec{q}\; (\vec{\bar{q}}) \rightarrow Q \bar{Q} q\; (\bar{q})$.
In the calculation of (i)-(iii) one encounters UV, IR, and mass (M)
singularities which are removed by renormalization, in the sum of
(i) and (ii), and by factorization, respectively \cite{bs}. To make all these
singularities manifest we choose to work in the framework of 
$n$-dimensional regularization.

The required polarized squared matrix elements 
for (\ref{eq:lopgf}) and (i)-(iii) 
\begin{equation}
\label{eq:polme}
\Delta\left|M\right|^2 = \frac{1}{2} 
\left[ \left|M\right|^2(++) - \left|M\right|^2(+-)\right] \;\;\;,
\end{equation}
where the $\pm$ denote the helicities of the incoming particles,
are obtained by projecting onto the helicity states of the bosons 
(photons or gluons) and quarks using the $\epsilon_{\mu\nu\rho\sigma}$ 
tensor and the $\gamma_5$ matrix, respectively \cite{bs}. 
By taking the sum instead of the difference in (\ref{eq:polme}) 
we fully agree with the known unpolarized results \cite{svn,ellis} 
and the abelian, $\vec{\gamma}\vec{\gamma}\rightarrow Q\bar{Q}$, part 
of our results agrees with \cite{conto} as well.

The presence of $\gamma_5$ and $\epsilon_{\mu\nu\rho\sigma}$ in the 
polarized calculation introduces some complications since these
objects have no unique continuation to $n\neq 4$. 
In the HVBM prescription, which we use, the usual $n$-dim.\ scalar 
products $k\cdot p$ are accompanied by their $(n-4)$-dim.\ subspace 
counterparts $\widehat{k\cdot p}$ (`hat momenta'). 
These terms deserve special attention when performing the 
$2\rightarrow 3$ phase space integrations.
For single-inclusive heavy quark production, which we consider here, 
only a single hat momenta combination $\hat{p}^2$ shows up in 
$\Delta\left|M\right|^2$ and the
appropriately modified $2\rightarrow 3$ phase space formula 
schematically reads \cite{bs}
\begin{equation}
\label{eq:dps3}
\mathrm{dPS}_3 = \mathrm{dPS}_{3,unp}(\theta_1,\theta_2) \times
\frac{1}{B\left(\frac{1}{2},\frac{n-4}{2}\right)} \int_0^1 dx\,
\frac{x^{(n-6)/2}}{\sqrt{1-x}}
\end{equation}
where $x\equiv 4 (s_4+m^2) \hat{p}^2/(s_4^2 \sin^2\theta_1 \sin^2\theta_2)$,
$B$ is the Beta function, $m$ denotes the heavy quark mass, and 
$s_4\equiv s+t_1+u_1$. 
$\theta_{1,2}$ are introduced to parametrize the momenta of the two 
not observed partons \cite{bs}. 
However, due to the appearance of $m$ in $x$ it turns out \cite{bs} that
{\em{all}} contributions due to $\hat{p}^2$ are at least of 
${\cal{O}}(n-4)$ and hence drop out when the limit $n \rightarrow 4$ is 
taken\footnote{This differs from a calculation involving only {\em massless}
particles.}. 
The remaining phase space integration \cite{bs} then proceeds as in the 
unpolarized case \cite{svn}.  

Finally, it should be recalled that in the 
factorization procedure for (iii) one has to introduce 
the parton content of the polarized {\em{photon}} \cite{photon} 
which is experimentally completely unknown so far. 
A scheme independent result in ${\cal{O}}(\alpha_s^2\alpha)$ 
can thus only be obtained for the sum of the `direct' {\em{and}} `resolved'
photon contributions. The NLO corrections for the latter
are unknown and, for the time being, have to be estimated in LO
\cite{svhera}.
\vspace*{-0.3cm}

\section{Numerical Results and Phenomenological Aspects}

The total photon-parton cross section in NLO can be expressed in terms of
scaling functions $(i=g,\,q,\,\bar{q})$
\begin{equation}
\Delta \hat{\sigma}_{i\gamma} (s,m^2,\mu_f)=
\frac{\alpha\alpha_s}{m^2} \left[ 
\Delta f_{i\gamma}^{(0)}(\eta) + 4\pi \alpha_s 
\left\{\Delta f_{i\gamma}^{(1)}(\eta) +
\Delta \bar{f}_{i\gamma}^{(1)}(\eta) \ln \frac{\mu_f^2}{m^2}\right\}
\right] \!\!
\label{eq:partonxsec}
\end{equation}
where $\Delta f_{i\gamma}^{(0)}$ and $\Delta f_{i\gamma}^{(1)}$,
$\Delta \bar{f}_{i\gamma}^{(1)}$ stand for the LO and NLO corrections,
respectively, $\mu_f$ denotes the factorization scale
(for simplicity we choose $\mu_r =\mu_f$), and $\eta\equiv s/4m^2 -1$. 
The scaling functions can be further decomposed depending on the
electric charge of the heavy and light quarks, $e_Q$ and $e_q$,
respectively:
\begin{equation}
\Delta f_{g\gamma} (\eta) = e_Q^2\, \Delta c_{g\gamma}(\eta)\;,\;\;
\Delta f_{q\gamma} (\eta) = e_Q^2\, \Delta c_{q\gamma} (\eta) +
e_q^2\, \Delta d_{q\gamma} (\eta)
\end{equation}
with corresponding expressions for the $\Delta \bar{f}_{i\gamma}$.

\begin{figure}[t]
\begin{center}
\vspace*{-1.15cm}
\epsfig{figure=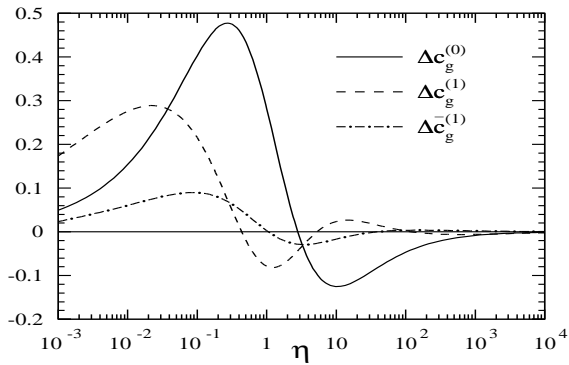,width=9cm,height=6cm} 
\vspace*{-0.7cm}
\caption{ The LO and NLO ($\overline{\mathrm{MS}}$)
gluonic scaling functions as a function of $\eta=s/4m^2-1$.}
\vspace*{-0.5cm}
\end{center}
\end{figure}
In Fig.~1 we present $\Delta c_{g\gamma}^{(0)}$, $\Delta c_{g\gamma}^{(1)}$,
and $\Delta \bar{c}_{g\gamma}^{(1)}$ as a function of $\eta$ 
in the $\overline{\mathrm{MS}}$ scheme. 
For the discussion below it is important to notice that
$\Delta c_{g\gamma}^{(0)}$ (solid line) changes sign at $\eta\simeq 3$. 
Upon adding the NLO terms, multiplied by a factor $4\pi\alpha_s$ 
(see (\ref{eq:partonxsec})), the zero is shifted towards $\eta\simeq 1$. 
We also note that for $\eta\lesssim 0.1$ the ${\cal{O}}(\alpha_s)$
corrections dominate over the LO result when we include that factor.
Comparing our polarized results with the unpolarized ones 
(see Fig.~5 in \cite{svn}), one observes that for $\eta \rightarrow 0$ 
$\Delta c_{g\gamma}\rightarrow c_{g\gamma}$, implying that
$\left|M_{g\gamma}\right|^2(+-) \rightarrow 0$.
On the contrary, for $\eta\rightarrow \infty$ 
the unpolarized NLO coefficients approach a plateau value \cite{svn}
dominating over the LO result while all polarized coefficients 
tend to zero here, implying that 
$\left|M_{g\gamma}\right|^2(++)\rightarrow \left|M_{g\gamma}\right|^2(+-)$.
The numerically less important quark coefficients can be found in \cite{bs}.

\begin{figure}[t]
\begin{center}
\vspace*{-1.15cm}
\epsfig{figure=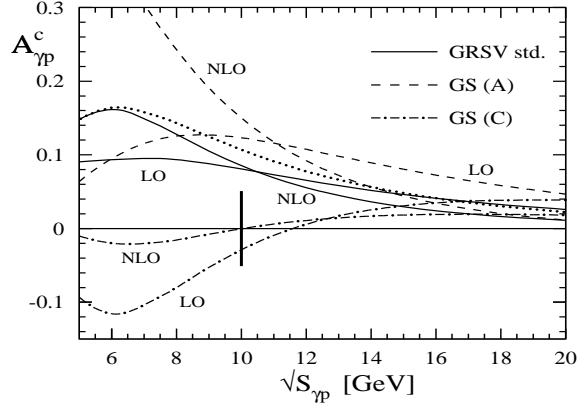,width=9cm, height=6.5cm}
\vspace*{-0.6cm}
\caption{$A_{\gamma p}^c$ in LO and NLO for
three sets of polarized parton densities $^{1,2}$. 
$\sigma_{\gamma p}^c$ in (6) was calculated using the GRV $^{11}$ densities. 
The dotted line shows $A_{\gamma p}^c$ in LO using a NLO $\Delta g$ (see text).
The bar shows the uncertainty for such a measurement at COMPASS $^{12}$.}
\vspace*{-0.7cm}
\end{center}
\end{figure}
Using (\ref{eq:partonxsec}) it is easy 
to calculate the experimentally relevant spin
asymmetry for the total hadronic heavy flavor photoproduction cross section 
\begin{equation}
\label{eq:asym}
A_{\gamma p}^Q(S_{\gamma p},m^2,\mu_f) = 
\Delta \sigma_{\gamma p}^Q (S_{\gamma p},m^2,\mu_f) /
\sigma_{\gamma p}^Q(S_{\gamma p},m^2,\mu_f)
\end{equation}
as a function of the photon-proton c.m.s.\ energy $S_{\gamma p}$ and
where
\begin{equation}
\label{eq:xsec}
\Delta \sigma_{\gamma p}^Q(S_{\gamma p},m^2,\mu_f) = 
\sum_{f=q,\bar{q},g}\int\limits_{4m^2/S_{\gamma p}}^1 dx\, 
\Delta \hat{\sigma}_{f \gamma}(x S_{\gamma p},m^2,\mu_f)\,
\Delta f^p(x,\mu_f^2)  
\end{equation}
(with a similar expression for the unpolarized cross section
$\sigma_{\gamma p}^Q$).

In Fig.~2 we show the charm asymmetry
$A_{\gamma p}^c$ in LO and NLO, using $m=1.5\;\mathrm{GeV}$, $\mu_f=2m$, and
three sets of polarized densities \cite{grsv,gs}, 
in the energy region relevant for COMPASS \cite{compass}
(they will operate at $\approx 10$ GeV). The NLO corrections are large, 
depend strongly on $\sqrt{S_{\gamma p}}$ and do not cancel in the 
ratio (\ref{eq:asym}) as one may naively expect. However, their origin
is readily explained \cite{bs}:
For GRSV \cite{grsv} and GS (A) \cite{gs} and 
$\sqrt{S_{\gamma p}}\gtrsim 12\,\mathrm{GeV}$ the corrections
stem from the shift of the zero in the gluonic coefficient function. 
$A_{\gamma p}^c$ changes sign at some 
$ \sqrt{S_{\gamma p}}> 20\,\mathrm{GeV}$ and large NLO corrections
in the vicinity of a zero are natural. 
For $\sqrt{S_{\gamma p}}\lesssim 12\,\mathrm{GeV}$, where one probes
$\Delta g$ at $x\gtrsim 0.1$, the corrections are, on the other hand, entirely
due to the badly constrained $\Delta g$, 
more precisely, due to the large differences between the LO and NLO $\Delta g$
in the two sets \cite{grsv,gs}. This is illustrated by the dotted curve
where we use the NLO GRSV gluon \cite{grsv} 
to calculate the LO $A_{\gamma p}^c$. 
In this energy region the observed large corrections thus 
should not be taken too literally.
For the GS (C) set \cite{gs} the situation is more involved since here
$\Delta g$ oscillates as well.

%
\begin{figure}[t]
\begin{center}
\vspace*{-1.15cm}
\epsfig{figure=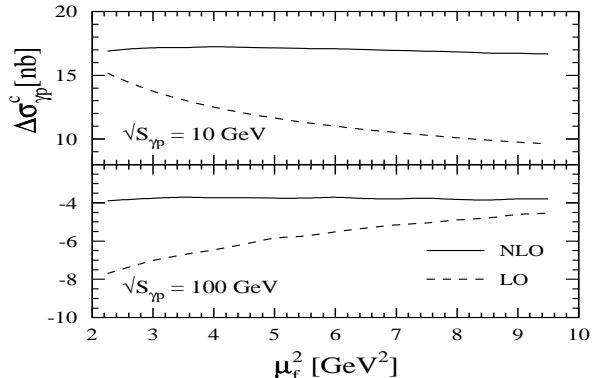,width=9cm, height=6cm}
\vspace*{-0.6cm}
\caption{The factorization scale dependence for the LO and NLO
$\Delta \sigma_{\gamma p}^c$ as defined in (7)
using the LO and NLO GRSV densities $^1$, respectively.}
\vspace*{-0.65cm}
\end{center}
\end{figure}
Finally we briefly discuss the importance of the main theoretical 
uncertainties.
Fig.~3 shows the dependence of $\Delta \sigma_{\gamma p}^c$ on the
choice of $\mu_f$. The improved scale dependence in NLO clearly
underlines the importance of our NLO calculation \cite{bs}. 
Light quark induced subprocesses contribute about $5\%$ at 
$\sqrt{S_{\gamma p}} \simeq 10\,\mathrm{GeV}$, but more for GS (C),
and should be subtracted before extracting $\Delta g$. 
The `resolved' photon contribution was shown to be small in \cite{svhera}.
More important is the unknown value of $m$, 
leading to shifts of about $30\%$ when $m$ is varied by 0.2 GeV
around 1.5 GeV used in our calculations.
Finally, for a reliable extraction of $\Delta g$ 
from $A_{\gamma p}^c$ our knowledge of the {\em{unpolarized}} gluon has to be 
improved as well since at large $x$ the uncertainty in $g(x,\mu^2)$ 
is rather sizeable \cite{gluon}. 
\vspace*{-0.25cm}

\section*{Acknowledgments}
It is a pleasure to thank I.\ Bojak for a fruitful collaboration.


\vspace*{-0.25cm}

\section*{References}
\begin{thebibliography}{99}
%
\bibitem{grsv} M.\ Gl\"uck {\it et al.}, \Journal{\PRD}{53}{4775}{1996}.
%
\bibitem{gs} T.\ Gehrmann and W.J.\ Stirling, \Journal{\PRD}{53}{6100}{1996}.
%
\bibitem{pdfs} G.\ Altarelli {\it et al.}, \Journal{\NPB}{496}{337}{1997},
{\tt hep-ph/9803237}; D.\ de Florian {\it et al.}, 
\Journal{\PRD}{57}{5803}{1998};
K.\ Abe {\it et al.}, E154 collab., \Journal{\PLB}{405}{180}{1997};
D.\ Adams {\it et al.}, SMC, \Journal{\PRD}{56}{5330}{1997}.
%
\bibitem{lohq} M.\ Gl\"uck and E.\ Reya, \Journal{\ZPC}{39}{569}{1988}.
%
\bibitem{svn} J.\ Smith and W.L.\ van Neerven, \Journal{\NPB}{374}{36}{1992}.
%
\bibitem{ellis} R.K.\ Ellis and P.\ Nason, \Journal{\NPB}{312}{551}{1989}.
%
\bibitem{bs} I.\ Bojak and M.\ Stratmann, {\tt hep-ph/9804353} and 
paper in preparation.
%
\bibitem{conto} B.\ Kamal {\it et al.}, \Journal{\PRD}{51}{4808}{1995}, 
D {\bf 55}, 3229(E) (1997);
G.\ Jikia and A.\ Tkabladze, \Journal{\PRD}{54}{2030}{1996}.
%
\bibitem{photon} M.\ Gl\"uck and W.\ Vogelsang, \Journal{\ZPC}{55}{353}{1992},
C {\bf 57}, 309 (1993);
M.\ Gl\"uck {\it et al.}, \Journal{\PLB}{337}{373}{1994};
M.\ Stratmann and W.\ Vogelsang, \Journal{\PLB}{386}{370}{1996}.
%
\bibitem{svhera}  M.\ Stratmann and W.\ Vogelsang, 
\Journal{\ZPC}{74}{641}{1997}. 
%
\bibitem{grv} M.\ Gl\"uck, E.\ Reya, and A.\ Vogt, 
\Journal{\ZPC}{67}{433}{1995}.
%
\bibitem{compass} G.\ Baum et al., COMPASS collab., CERN/SPSLC-96-14.
%
\bibitem{gluon} W.\ Vogelsang and A.\ Vogt, \Journal{\NPB}{453}{334}{1995};
J.\ Huston {\it et al.}, {\tt hep-ph/9801444};
A.D.\ Martin {\it et al.}, {\tt hep-ph/9803445}.
%
\end{thebibliography}
\end{document}